\documentclass[12pt]{article}
\newcommand{\be}{\begin{equation}}
\newcommand{\ee}{\end{equation}}
\newcommand{\bea}{\begin{eqnarray}}
\newcommand{\eea}{\end{eqnarray}}
\newcommand{\nn}{\nonumber \\}

\def\({\left(}
\def\){\right)}
\def\a{\alpha}
\def\b{\beta}

\begin{document}

\begin{center}
{\Large\bf Exact ${\cal N}=4$ Supersymmetric
Low-Energy Effective Action in ${\cal N}=4$
  Super Yang-Mills Theory}

{I.L.~Buchbinder}

{\it Department of Theoretical Physics, Tomsk State Pedagogical
University,\\ 
Tomsk 634041, Russia} 

{E.A.~Ivanov}

{\it Bogoliubov Laboratory of Theoretical Physics, Joint Institute for
  Nuclear Research,\\ Dubna, 141980 Moscow Region, Russia} 

\end{center}


\begin{abstract}
  We review a recent progress in constructing the low-energy effective
  action in ${\cal N}=4$ SYM theory. This theory is formulated in
  terms of ${\cal N}=2$ harmonic superfields corresponding to 
  ${\cal N}=2$ vector multiplet and hypetrmultiplet. Such a formulation
  possesses the manifest ${\cal N}=2$ supersymmetry and an extra
  hidden on-shell supersymmetry. Exploring this hidden ${\cal N}=2$
  supersymmetry we prove that the known non-holomorphic potentials of
  the form $\ln W \ln \bar W$ can be explicitly completed by the
  appropriate hypermultiplet-dependent terms to the entire  
  ${\cal N}=4$ supersymmetric form.  The non-logarithmic effective
  potentials do not admit an ${\cal N}=4$ completion and, hence, such
  potentials cannot occur in ${\cal N}=4$ supersymmetric theory.
  As a result, we obtain the exact ${\cal N}=4$ supersymmetric
  low-energy effective action in ${\cal N}=4$ SYM theory
\end{abstract}

\section{Introduction}

The study of various aspects of the ${\cal N}=4$ supersymmetric Yang-Mills
(SYM)
theory is one of the most active trends in modern high energy
theoretical 
physics. We will discuss here a recent progress towards 
finding out
the ${\cal N}=4$ SYM low-energy effective action. \footnote{ By the 
low-energy effective action we always mean the leading in the external 
momenta piece of the full quantum effective action.}

First consideration of the low-energy effective action in ${\cal N}=2$
superconformal theories including ${\cal N}=4$ SYM theory has been
given 
in \cite{WGR, DS}. In ${\cal N}=4$ SYM model with gauge group $SU(2)$ 
spontaneously broken down to $U(1)$ (Coulomb branch), the requirements
of 
the scale and $R$-invariance determine the part of the effective
action 
depending on ${\cal N}=2$ superfield strengths $W$ and $\bar W$ up to
a 
numerical coefficient. The result is formulated in terms of
non-holomorphic 
effective potential
\begin{equation}
{\cal
H}(W,\bar{W})=c\,\ln\frac{W}{\Lambda}\,\ln\frac{\bar{W}}{\Lambda}~.
\label{1}
\end{equation}
Here $\Lambda$ is an arbitrary scale and $c$ is an arbitrary real
coefficient.
The effective action in the ${\cal N}=2$ gauge fields sector is defined as
an
integral 
of ${\cal H}(W,\bar{W})$ over the full ${\cal N}=2$ superspace. One
can show
\cite{DS,LU} that the potential (\ref{1}) gets neither perturbative 
contributions beyond one loop nor 
non-perturbative corrections. As a result, expression (\ref{1}) 
determines the exact low-energy effective action for the $SU(2)$ 
theory in the Coulomb branch. The problem is reduced to computing the 
coefficient $c$ in one-loop approximation.

The direct derivation of the potential (\ref{1}), computation of the 
coefficient $c$
and, hence, the final reconstruction of the exact exact low-energy
$U(1)$ 
effective action
from the quantum ${\cal N}=4$ SYM theory were given in refs. 
\cite{PU}-\cite{BK1}. Later these results were generalized to the group
$SU(N)$
broken to its
maximal abelian subgroup \cite{CT}-\cite{BBK}, \cite{LU}. 
The corresponding one-loop effective
potential is given by
\begin{equation}
{\cal H}(W,\bar{W})=c\sum\limits_{I<J}\ln\frac{W^I-W^J}{\Lambda}\,
\ln\frac{\bar{W}^I-\bar{W}^J}{\Lambda}~,  
\label{2}
\end{equation}
with the same coefficient $c$ as for the group $SU(2)$. Here
$I,J=1,2,\dots,N$, $W=\sum\limits_1W^Ie_{II}$ belongs to Cartan
subalgebra of the algebra $su(N)$, $\sum\limits_iW^I=0$, and $e_{IJ}$ is
the Weyl basis in the $su(N)$ algebra.

The potential (\ref{2}) looks quite analogous to (\ref {1}). However, 
we cannot
state that (\ref{2}) determines the exact low-energy effective action
as in 
the case of $SU(2)$ group.
The general scale and R-invariance considerations do not forbid
the presence of some extra terms in the non-holomorphic
effective potential of the form \cite{LU}, \cite{DG}
\begin{equation}
f\left(\frac{W^I-W^J}{W^K-W^L}\,,
\frac{\bar{W}^I-\bar{W}^J} {\bar{W}^K-\bar{W}^L}\right)~,  \label{3}
\end{equation}
with $f$ being real functions. 
The
direct calculations \cite{BP} have not confirmed the appearance of 
terms like (\ref{3}) 
at two, three and four loops. However, in a general setting, the
problem
of contributions (\ref{3}) to the effective action remained open.

We wish to pay attention to the fact that all the results concerning the 
structure
of the low-energy effective action of ${\cal N}=4$ SYM theory were
actually 
obtained only for a part of it, that defined  in the ${\cal N}=2$ gauge
fields sector.
From the point of view of ${\cal N}=2$ supersymmetry, the ${\cal N}=4$ 
gauge multiplet consists of ${\cal N}=2$
vector multiplet and hypermultiplet (see e.g. \cite{harm2}). The
problem
of the hypermultiplet dependence of ${\cal N}=4$ SYM effective action 
has been studied in our papers \cite{BI, BIP} where the exact 
${\cal N}=4$ low-energy effective action has been constructed. 
The aim of
the 
present paper is to give a brief review of the approach 
of Ref.
\cite{BI}.

A natural framework for finding the ${\cal N}=4$ supersymmetric effective
action 
is provided by a formulation of ${\cal N}=4$ SYM theory in terms of
superfields
carrying off-shell representations of ${\cal N}=2$ supersymmetry 
defined in ${\cal N}=2$ harmonic superspace \cite{harm1},
\cite{harm2}. 
The harmonic 
superspace approach was succesfully used to study the effective action 
in extended supersymmetric theories in refs. 
\cite{BK1}, \cite{BBK}, \cite{BP}-\cite{KM}
(see also the review \cite{BBIKO}).

To find the ${\cal N}=4$ supersymmetric effective action, we 
proceed in the following way. We begin with the ${\cal N}=4$ SYM theory
formulated in terms of ${\cal N}=2$ harmonic superfields. 
Then we examine which hypermultiplet-dependent terms can be added 
to the potentials (\ref{1}) -
(\ref{3}) in order to ensure full ${\cal N}=4$ supersymmetry of the 
effective action. 
We find such terms for the potentials 
(\ref{1}) and (\ref{2}) and show that the analogous terms do not exist
for 
the potential (\ref{3}) \cite{BI}. Therefore, the potentials of the
form 
(\ref{3}) can never occur in the full supersymmetric theory.

\section{${\cal N}=4$ SYM Theory in ${\cal N}=2$ Harmonic Superspace}

The most efficient approach to constructing quantum 
formulations of supersymmetric field theory models is based on 
the use of superfields carrying off-shell representations
of supersymmetry (see e.g. \cite{BK}). The main attractive feature of
such 
an approach is the possibility to preserve a manifest supersymmetry 
on all steps  of quantum calculations. From this point of view, the 
most appropriate quantum formulation of ${\cal N}=4$ SYM theory 
would be the one in ${\cal N}=4$ superspace. However, the corresponding
formulation is unknown so far. The best what we can employ at present
for ${\cal N}=4$ SYM theory is its formulation in terms of ${\cal N}=2$
superfields in harmonic superspace \cite{harm2}, \cite{harm1}. 
In this case two supersymmetries are manifest and two other ones are hidden.

The action of ${\cal N}=4$ SYM theory is written in ${\cal N}=2$ 
harmonic superspace as a sum of actions for ${\cal N}=2$ gauge 
multiplet and hypermultiplet in the adjoint repesentation coupled to 
the gauge multiplet
\begin{eqnarray}
&& S[V^{++},q^+]=
\frac{1}{8}\left(\int d^8\zeta_L {\rm tr\, W^2}+
\int d^8\zeta_R {\rm tr\, \bar{W}^2}\right) \nonumber \\
&& -\,
\frac{1}{2}\int d\zeta^{(-4)}{\rm tr\,}q^{+a}
\left(D^{++}+igV^{++}\right)q^+_a~.
\label{4}
\end{eqnarray}
Here the real analytic superfield $V^{++}$ is an unconstrained 
gauge potential of ${\cal N}=2$ SYM theory and the unconstrained 
charged analytic superfield $q^{+}_{a}$, $a=1,2,$ describes the
hypermultiplet.
The action (\ref{4}) is manifestly ${\cal N}=2$ supersymmetric.
Moreover, this action possesses an extra hidden ${\cal N}=2$ supersymmetry
which mixes ${\cal N}=2$ superfield strengths $W, \bar W$ with $q^+_a$
\cite{harm2}. As a result, the action (\ref{4}) describes 
the ${\cal N}=4$ supersymmetric YM theory. Our aim is to construct 
${\cal N}=4$ supersymmetric effective action whose classical limit would be
the action (\ref{4}) and whose hypermultiplet-independent part 
is the known ${\cal N}=2$ supersymmetric functional corresponding to
the
low-energy effective action in the ${\cal N}=2$ gauge multiplet sector.
The latter 
action is given in terms of non-holomorphic effective
potential
${\cal H}(W,\bar{W})$ depending on the abelian chiral and antichiral 
superfield strengths $W$ and $\bar W$ satisfying the free classical
equations of motion
(on-shell conditions)
\begin{equation}
(D^+)^2 W = 0, \quad(\bar D^+)^2 \bar W = 0~, 
\label{5}
\end{equation}
where
$(D^+)^2=D^{+\alpha}D^+_{\alpha}$, 
$(\bar D^+)^2=\bar D^{+}_{\dot\alpha}\bar D^{+\dot\alpha}$,
$D^{\pm}_{\alpha}=u^{\pm}_i D^{i}_{\alpha}$,
$D^{\pm}_{\dot\alpha}=u^{\pm}_i \bar D^{i}_{\dot\alpha}$
and $D^{i}_{\alpha}, \bar D^{i}_{\dot\alpha}$ are the standard 
${\cal N}=2$ supercovariant derivatives. As a result, we have to know
the 
hidden ${\cal N}=2$ supersymmetry transformations only 
on the free 
mass shell
(\ref{5}).
Obviously, the corresponding hypermultiplet should also be assumed 
to satisfy the free classical equations of motion
\begin{equation}
D^{++}q^{+}_a =0.
\label{6}
\end{equation}
Taking into account eqs. (\ref{5}, (\ref{6}), one can write the
hidden 
${\cal N}=2$ supersymmetry transformations in the form \cite{harm2}
\begin{eqnarray}
&& \delta W = {1\over 2}\bar\epsilon^{\dot\alpha a}\,\bar
D^-_{\dot\alpha}q^+_a\;, \quad
\delta \bar W = {1\over 2}\epsilon^{\alpha a}\,
D^-_{\alpha}q^+_a~, \nn
&& \delta q^+_a ={1\over 4}\,(\epsilon^\beta_a D^+_\beta W +
\bar\epsilon^{\dot\alpha}_a\bar D^+_{\dot\alpha} \bar W)~, \quad
\delta q^-_a ={1\over 4}\,(\epsilon^\beta_a D^-_\beta W +
\bar\epsilon^{\dot\alpha}_a\bar D^-_{\dot\alpha} \bar W)~, 
\label{7}     
\end{eqnarray}
where $\epsilon^{\alpha a}, \bar \epsilon^{\dot\alpha a}$ are  
anticommuting infinitesimal parameters. 
For  further use, we have introduced the quantity 
\begin{equation}
q^{-a}=D^{--}q^{+a}
\label{8}
\end{equation}
satisfying the on-shell relations
\begin{equation}
D^{--}q^{-a} =0, \quad D^{++}q^{-a} =q^{+a}, 
D_{\alpha}^- q^{-a}=\bar D^{-}_{\dot\alpha}q^{-a}=0.
\label{9}
\end{equation}
The operators $D^{++}, D^{--}$ are given in \cite{harm2}. 

We would like to emphasize once more that the supersymmetry 
transformations (\ref{7}) are essentially on-shell.

\section{Construction of {\cal N}=4 Supersymmetric Effective \break 
Action}

We begin with the $SU(2)$ gauge theory. The effective action is assumed to
have
the following general form 
\begin{equation}
\Gamma[W, \bar W, q^+]= S[V^{++}, q^+]+\bar \Gamma[W, \bar W, q^+],
\label{11}
\end{equation}
where $S[V^{++}, q^+]$ is the classical action (\ref{4}) and the 
functional $\bar \Gamma[W, \bar W, q^+]$ incorporates quantum
corrections. 
We also assume that the functional $\bar \Gamma[W, \bar W, q^+]$
satisfies the condition
\begin{equation}
\bar \Gamma[W, \bar W, q^+]|_{q^{+}=0}=\int d^{12}zdu
{\cal H}(W,\bar{W})~,
\label{12}
\end{equation}
with ${\cal H}(W,\bar{W})$ being the known non-holomorphic potential.
Of course, the integral over harmonics is equal to unity, we have
written
it for further convinience. Here $d^{12}z$ is the full ${\cal N}=2$ 
superspace measure. Eqs. (\ref{11}), (\ref{12}) imply that the 
corrections to the effective action are of the form 
\begin{eqnarray}
\Gamma[W,\bar{W},q^+] &=& \int d^{12}zdu
\left[{\cal H}(W,\bar{W})+
{\cal L}_q(W,\bar{W},q^+)\right] \nonumber \\
&=&  \int d^{12}zdu\,
{\cal L}_{eff}
(W,\bar W, q^+)~. \label{13}
\end{eqnarray}
${\cal L}_q(W,\bar{W},q^+)$ is some function unknown for a moment.
The functional (\ref{13}) is manifestly ${\cal N}=2$ supersymmetric.
We will seek for a function ${\cal L}_q(W,\bar{W},q^+)$ such that 
the full functional is invariant under the hidden ${\cal N}=2$
supersymmetry transformations (\ref{7}).

Let us consider the transformation of the first term in (\ref{13})
under (\ref{7}), taking into account the explicit 
form of ${\cal H}(W,\bar{W})$ (\ref{1}). One obtains
\begin{equation}
\delta \int d^{12}zdu\, {\cal H} (W, \bar W) = \frac{1}{2} c \int
d^{12}zdu \frac{q^{+ a}}{\bar W W} ( \epsilon^\alpha_{a}
D^-_\alpha W + \bar \epsilon^{\dot \alpha}_{a} \bar D^-_{\dot
\alpha} \bar W )~. 
\label{14}
\end{equation}
The function ${\cal L}_q (W, \bar W, q^+)$ 
is determined from the condition that
the variation of the second term in (\ref{13}) cancels
the variation (\ref{14}).
We will search for  ${\cal L}_q (W, \bar W, q^+)$ in the form of the
series
\cite{BI}
\begin{equation}
{\cal L}_q = c
\sum^{\infty}_{n=1} c_n \left( \frac{q^{+a}q^{-}_a }{\bar W W}
\right)^n 
\label{15}
\end{equation}
with some unknown coefficients $c_n$. The quantity $q^{-}_{a}$ is
defined 
by (\ref{8}). The condition 
\begin{equation}
c
\sum^{\infty}_{n=1} c_n 
\delta \int d^{12}z du
\left( \frac{q^{+a}q^{-}_a }{\bar W W} \right)^n
=-\delta \int d^{12}zdu\, {\cal H} (W, \bar W)
\label{16}
\end{equation}
where the variation of the right hand side is given by (\ref{14}),
allows us to obtain the recursive relations between the coefficients 
$c_{n-1}$ and $c_n$ (see \cite{BI} for details)
\be
c_n = - 2 \frac{(n-1)^2}{n(n+1)} c_{n-1}, \quad c_1=-1.  
\label{17}
\end{equation}
The relations (\ref{17}) are solved by
\begin{equation}
c_n = \frac{(-2)^n}{n^2(n+1)}~.
\label{17.5}
\end{equation}
As the result, the function ${\cal L}_q$ proves to be
\begin{equation}
{\cal L}_q (W, \bar W, q^+) 
\equiv  {\cal L}_q (X) = c\, \sum_{n=1}^{\infty} \frac{1}{n^2
(n+1)} X^n, 
\label{18}
\end{equation}
where
\begin{equation} 
X= -2\, \frac{q^{+a}q^{-}_a
}{\bar W W} \label{19}.
\end{equation}
The series in the right hand side of (\ref{18}) can be 
rewritten in terms of Euler dilogarithm function $\mbox{Li}_2(X)$
\cite{BE},
\begin{equation}
\mbox{Li}_2(X)=\sum_{n=1}^{\infty} \frac{X^n}{n^2}=-\int_0^X
\frac{\mbox{ln}(1-t)}{t}dt,
\label{20}
\end{equation}
as
\begin{equation}
{\cal L}_q (W, \bar W, q^+)=c \left( (X-1)\frac{\mbox{ln}(X-1)}{X} +
[\mbox{Li}_2(X)-1] \right).
\label{21}
\end{equation}
The quantity $X$ defined in (\ref{19}) actually does not depend on
harmonics
since the quantities $W, \bar W$ and $q^{+a}, q^{-}_{a}$ are on-shell 
and obey eqs. (\ref{5}), (\ref{6}), (\ref{9}):
\begin{equation}
X=-\frac{q^{ia}q_{ia}}{\bar W W}.
\label{22}
\end{equation}
Therefore ${\cal L}_q (W, \bar W, q^+)$ does not depend on harmonics 
and the integral over harmonics in (\ref{13}) can be omitted. As the
result,
the full ${\cal N}=4$ supersymmetric low-energy effective action has
the form 
\be
\Gamma [ W, \bar W, q^+] = S[V^{++}, q^{+}]+ \int d^{12}z\,
 {\cal L}_{eff} (W, \bar W, q^+)~, 
\label{23}
\ee
where
\be
{\cal L}_{eff} (W, \bar W, q^+) = {\cal H} (W, \bar W) +
{\cal L}_q (X).  
\label{24}
\ee
Here ${\cal H} (W, \bar W)$ is given by eq. (\ref{1}),
${\cal L}_q (X)$ by eq. (\ref{21}) and $X$ by eq. (\ref{22}).

We emphasize that $\Gamma[W, \bar W, q^+]$ (\ref{23}) is the exact 
low-energy effective action. It is known that the non-holomorphic 
effective potential ${\cal H} (W, \bar W)$ (\ref{1}) is exact \cite{DS}.
The function ${\cal L}_q (X)$ (\ref{21}) was uniquely restored from
(\ref{1})
by invoking the requirement of the 
hidden ${\cal N}=2$ supersymmetry (\ref{7}). 
Hence, it is the only function which forms, together with
${\cal H} (W, \bar W)$, the ${\cal N}=4$ supersymmetric functional 
(\ref{23}), (\ref{24}). Therefore, the functional (\ref{23}), (\ref{24})
is the exact low-energy effective action of  ${\cal N}=4$ SYM
theory with the $SU(2)$ gauge group spontaneously broken down 
to $U(1)$.

We have obtained the effective Lagrangian (\ref{24}), (\ref{21}) 
from the purely algebraic considerations. Recently, we have shown that
this effective Lagrangian can be recovered in the framework of 
quantum field theory by calculating the one-loop harmonic supergraphs 
with an arbitrary number of the external hypermultiplet legs \cite{BIP}.

\section{Component Structure and Generalization to 
$SU(N)$ Gauge Group}

We consider the component form of the effective action (\ref{23})
in the bosonic sector. In this case
\begin{eqnarray}
&&W = \varphi(x) + 4 i \theta^+_{(\a} \theta^-_{\b )} F^{(\a\b)}(x)~,
\quad
\bar W = \bar \varphi(x) + 
4 i \bar \theta^+_{(\dot\a} \bar \theta^-_{\dot\b )}(x)
\bar F^{(\dot\a\dot\b)}(x)~, 
\nonumber \\
&& D^+_\a D^-_\b W = -4i F_{(\a\b)}~, 
\quad \bar D^+_{\dot\a} \bar D^-_{\dot\b} \bar W =
4i \bar F_{(\dot\a\dot\b)}~, \quad q^{ia} = f^{ia}(x)~.
\label{25}
\end{eqnarray}
Here $\varphi(x)$ is a complex scalar field 
belonging to the ${\cal N}=2$ vector multiplet, $F^{\alpha\beta}(x)$ and
$\bar
F^{\dot{\alpha}\dot{\beta}}(x)$ are the self-dual and anti-self-dual
components of the abelian field strength $F_{mn}$, and $f^{ia}(x)$
stands for four scalar fields of the hypermultiplet.
The quantity $X$ in the bosonic sector is reduced to 
\begin{equation}
X |_{\theta = 0} = - \frac{f^{ia}f_{ia}}{|\varphi|^2} \equiv X_0~.
\label{25.5}
\end{equation}
To find the structure of the low-energy effective action in the
bosonic 
sector, one uses the relation 
\begin{equation}
\int d^{12}z\,
{\cal L}_{eff} = \frac{1}{16^2} \int d^4 x\, (D^+)^2 (D^-)^2 (\bar
D^+)^2
(\bar D^-)^2 {\cal L}_{eff}. 
\label{26}
\end{equation}
Fulfilling the differentiation in (\ref{26}) and discarding fermions, one
gets 
\begin{equation}
\int d^{12}z\,
{\cal L}_{eff} = 
4c\,
\int d^4 x \frac{F^2 \bar F^2}{|\varphi |^4} \left[1 + G(X_0
)\right]~, 
\label{27}
\end{equation}
where \cite{BI}
\begin{equation}
G(X_0 ) =\frac{X_0\, (2-X_0)}{(1-X_0 )^2}~.
\label{28}
\end{equation}
Here  
$F^2 = F^{\alpha\beta}F_{\alpha\beta},
 \bar F^2 = \bar F^{\dot{\alpha}\dot{\beta}}\bar
F_{\dot{\alpha}\dot{\beta}}$. 
Now one substitutes the explicit form of $X_0$ (\ref{25.5})
into (\ref{27}), (\ref{28})
and finally obtains the quantum correction 
functional $\bar \Gamma$ in the bosonic sector in the extremely simple form 
\begin{equation}
\bar \Gamma^{bos} = 4c \int d^4 x
\frac{F^2 \bar F^2}{(| \varphi|^2 + f^{ia}f_{ia})^2}~. \label{29}
\end{equation}
We point out that the denominator here is $SU(4)$ invariant square
of six real fields of the ${\cal N}=4$ vector multiplet.

Now let us turn to the theory with the gauge group $SU(N)$
spontaneously
broken down to its maximal abelian subgroup $U(1)^{N-1}$. In this case
the 
non-holomorphic effective potential is given by (\ref{2}). Its 
structure is analogous to (\ref{1}) and, therefore, we can again apply
the 
techniques exposed in Section 3. This leads to (see \cite{BI} for details)
\begin{equation}
{\cal L}_{eff}(W, \bar W, q^{+}) =\sum_{I,J} {\cal L}_{eff}^{IJ}
(W, \bar W, q^{+}) .
\label{30}
\end{equation}
Here, each contribution ${\cal L}_{eff}^{IJ}
(W, \bar W, q^{+})$ has the form (\ref{24}), (\ref{21}), with $X$ being
replaced 
by
\begin{equation}
X_{IJ}=-\frac{q^{ia}_{IJ}q_{iaIJ}}{W_{IJ}\bar W_{IJ}},
\label{31}
\end{equation}
$W_{IJ}=W_{I}-W_{J}, \bar W_{IJ}=\bar W_{I}-\bar W_{J}$, 
and $q^{ia}_{IJ}=q^{ia}_I=q^{ia}_J$.
The hypermultiplet superfields $q^{ia}$ belong to the Cartan 
subalgebra of the Lie algebra $su(N)$. The quantum correction
functional $\bar \Gamma[W, \bar W, q^{+}]$ in the bosonic
sector is given by the sum of terms (\ref{29}), as follows from
(\ref{30}).

Now we are going to examine if the non-holomorphic potential 
(\ref{3}) admits an ${\cal N}=4$ completion. The corresponding 
${\cal N}=4$ supersymmetric quantum correction functional must be of
the 
form
\begin{equation}
\int d^{12}z du \{ f(V_{IJKL}, \bar V_{IJKL}) +
{\cal L}_{q}(W_{IJ},W_{KL},\bar W_{IJ},\bar W_{KL},q^{+ a}_{IJ}, q^{+
a}_{KL})\}~, \label{32}
\end{equation}
where
\begin{equation}
V_{IJKL} = \frac{W_{IJ}}{W_{KL}}~,
\quad \bar V_{IJKL} = \frac{\bar W_{IJ}}{\bar W_{KL}}
\label{33}
\end{equation}
and
$W_{IJ}, \bar W_{IJ},
q^{+a}_{IJ}$ are given in (\ref{31}).
Here 
${\cal L}_{q}(W_{IJ},W_{KL},\bar W_{IJ},\bar W_{KL},q^{+ a}_{IJ}, q^{+
a}_{KL})\}$ 
is some
unknown function 
which must constitute, together with $f(V_{IJKL}, \bar V_{IJKL})$
(\ref{3}), the full ${\cal N}=4$ supersymmetric functional. 
The functional (\ref{32}) is manifestly ${\cal N}=2$ supersymmetric.
To prove its ${\cal N}=4$ supersymmetry we have to examine its
variation 
under the hidden ${\cal N}=2$ supersymmetry (\ref{7}). However, one
can show
(see \cite{BI} for details) that the condition of invariance under 
these transformations, even in the lowest order in the hypermultiplet 
superfields, leads to the meaningless relations 
\begin{equation}
\frac{W_{KL}}{W_{IJ}} = \frac{\bar W_{KL}}{\bar W_{IJ}}.
\label{33.5}
\end{equation}
This implies that the appropriate function ${\cal L}_q$ in 
(\ref{32}) does not exist. We thus conclude that the effective potential 
(\ref{3}) cannot appear in ${\cal N}=4$ SYM theory since
its ${\cal N}=4$ completion cannot be constructed. Clearly,
our consideration does not rule out a possibility 
of appearance of
the effective
potential (\ref{3}) in generic ${\cal N}=2$ superconformal models which 
do not possess ${\cal N}=4$ supersymmetry. Thus, the exact 
${\cal N}=4$ SYM theory with the gauge group $SU(N)$ spontaneously
broken down to $U(1)^{N-1}$ is given by eqs. (\ref{30}), (\ref{31}).

\section{Conclusions}

We have studied the problem of the low-energy effective action
depending on 
all fields of the ${\cal N}=4$ gauge multiplet of ${\cal N}=4$ SYM
theory in the Coulomb branch.
Using a formulation of ${\cal N}=4$ SYM theory in terms of ${\cal
N}=2$ 
harmonic superfields and exploring the hidden ${\cal N}=2$
supersymmetry 
of this theory we proved that the known non-holomorphic effective 
potentials (\ref{1}), (\ref{2}) can be uniquely completed to ${\cal
N}=4$ 
supersymmetric form by adding the appropriate hypermultiplet-dependent
terms and found the exact form of these terms. The same analysis,
being applied to the non-holomorphic effective potential (\ref{3}), 
showed that such a potential can never appear in the ${\cal N}=4$
supersymmetric theory. As a result, we have
found the exact ${\cal N}=4$ supersymmetirc effective action in the
Coulomb branch of ${\cal N}=4$ SYM theory (further details can be found in
\cite{BI, BIP}).

\section*{Acknowledgements} 

The work of I.L.B. and E.A.I. was partially supported by INTAS grant, 
project No 00-00254, DFG grant, project No 436 RUS 113/669 and
RFBR-DFG grant, project No 02-02-04002. The work of E.A.I.
was partially supported by RFBR-CNRS grant,
project No 01-02-22005.

\end{document}